\documentclass{article}

\usepackage{PRIMEarxiv}

\usepackage[utf8]{inputenc} 
\usepackage[T1]{fontenc}    
\usepackage{hyperref}       
\usepackage{url}            
\usepackage{booktabs}       
\usepackage{amsfonts}       
\usepackage{nicefrac}       
\usepackage{microtype}      
\usepackage{lipsum}
\usepackage{fancyhdr}       
\usepackage{graphicx}       
\graphicspath{{media/}}     
\usepackage{multirow}
\usepackage{cite}

\title{Evaluation of Deep Learning Topcoders Method for Neuron Individualization in Histological Macaque Brain Section
\thanks{\textit{\underline{Citation}}: 
\textbf{Wu, Huaqian, et al. "Evaluation of Deep Learning Topcoders Method for Neuron Individualization in Histological Macaque Brain Section." 2021 43rd Annual International Conference of the IEEE Engineering in Medicine \& Biology Society (EMBC). IEEE, 2021.}} 
}

\author{%
Huaqian Wu$^1$ \and
Nicolas Souedet$^1$ \and
Zhenzhen You$^2$ \and 
Caroline Jan$^1$ \and
\ C\'edric Clouchoux$^3$ and Thierry Delzescaux$^1$ \\
\\
$^1$CEA-CNRS-UMR 9199, MIRCen, Fontenay-aux-Roses, France
 \and
\ $^2$School of Computer Science and Engineering, Xi'an University of Technology, Xi'an, China.
 \and
\ $^3$WITSEE, Paris, France
}%

\begin{document}
\maketitle

\begin{abstract}
Cell individualization has a vital role in digital pathology image analysis. Deep Learning is considered as an efficient tool for instance segmentation tasks, including cell individualization. However, the precision of the Deep Learning model relies on massive unbiased dataset and manual pixel-level annotations, which is labor intensive. Moreover, most applications of Deep Learning have been developed for processing oncological data. To overcome these challenges, i) we established a pipeline to synthesize pixel-level labels with only point annotations provided; ii) we tested an ensemble Deep Learning algorithm to perform cell individualization on neurological data. Results suggest that the proposed method successfully segments neuronal cells in both object-level and pixel-level, with an average detection accuracy of 0.93.
\end{abstract}


\section{Introduction}
Lack of information about neuron population, distribution and morphology at cell level has existed as a critical problem for the study of brain development and aging for many years. Traditionally, neurobiologists estimate manually the number of neurons in the region of interest. However, this method is tedious and subjective because its accuracy relies on the complexity of images. Several automated cell individualization methods have been proposed such as Watershed~\cite{cousty2008watershed} and iCut~\cite{he2015icut}. However, these methods have several limitations. Watershed algorithm can be easily affected by noise in the images, often resulting in over- and under-segmentations. The iCut algorithm proposed to segment touching cells, fails in the regions where massive cells aggregate, and does not take into account size-varying cells such as neurons. Recently, the development of Deep Learning (DL) has revolutionized computer vision. Integrating DL models in cell detection and cell instance segmentation improves accuracy compared to traditional approaches~\cite{falk2019u, cui2019deep, naylor2018segmentation, kumar2017dataset}. To achieve robust and rigorous segmentation, a large training dataset is mandatory. However, pixel-level annotation is laborious and time consuming. In addition, aforementioned methods are designed mainly for analyzing oncological data, in particular H\&E staining, whereas in neuroscience, the study of neurons usually relies on NeuN staining. Moreover, the task of neuron segmentation is extremely challenging due to the variety in neuron shape, size and density in the brain.  To the best of our knowledge, few studies have specifically investigated cell individualization for neurons~\cite{you2019automated}.

In this paper, we propose a weakly supervised DL method for neuron instance segmentation, which requires only point annotations. The main contributions of this work are as follows: i) inspired by~\cite{you2019automated}, we developed a new strategy to synthesize pixel-level labels using Random Forest (RF) segmentation and a competitive Region Growing algorithm; ii) we tested different configurations of DL networks on NeuN stained images. The DL architecture used in this work is Topcoders, which ranked first during the nucleus segmentation competition: Data Science Bowl 2018 (DSB) and demonstrated promising performances on H\&E staining images and fluorescent images stained with DAPI and Hoechst~\cite{caicedo2019nucleus} and iii) by employing an overlapping patch extraction/assembling method [4], we were able to process large high-resolution images despite the limitation of GPU random access memory (RAM).

\section{Materials and Methods}
\subsection{Dataset}
\begin{figure}
  \centering
  \includegraphics[width=0.7\textwidth]{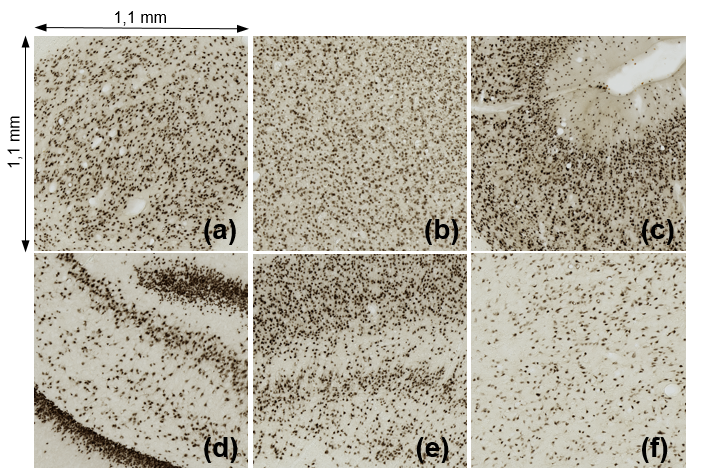}
  \caption{Training dataset examples: (a) caudate, (b-c) cortex, (d) hippocampus, (e) subiculum and (f) thalamus.}
  \label{fig:fig1}
\end{figure}

Dataset was derived from a previously published study~\cite{you2019automated}, a representative histological section with thickness of 40 $\mu$ m was obtained from a healthy macaque brain and scanned by an AxioScan.Z1 (Zeiss) with the resolution of 0.22 $\mu$ m/pixel ($\times$ 20 magnification) ($\sim$ 150 GB). Based on 30 images of 5000 $\times$ 5000 pixels, 24 images were chosen to create the training dataset including 11k patches (224 $\times$ 224 pixels), 6 other images were chosen as the test dataset. The datasets contained the main anatomical regions where the neuron distribution is investigated including cortex, hippocampus, caudate, etc. The datasets presented a rich diversity in terms of neuron shape and size, with both sparse and highly aggregated neuron distributions, as illustrated in Figure \ref{fig:fig1} .

\subsection{Pipeline of pixel-level label synthesis}
\begin{figure}
  \centering
  \includegraphics[width=0.7\textwidth]{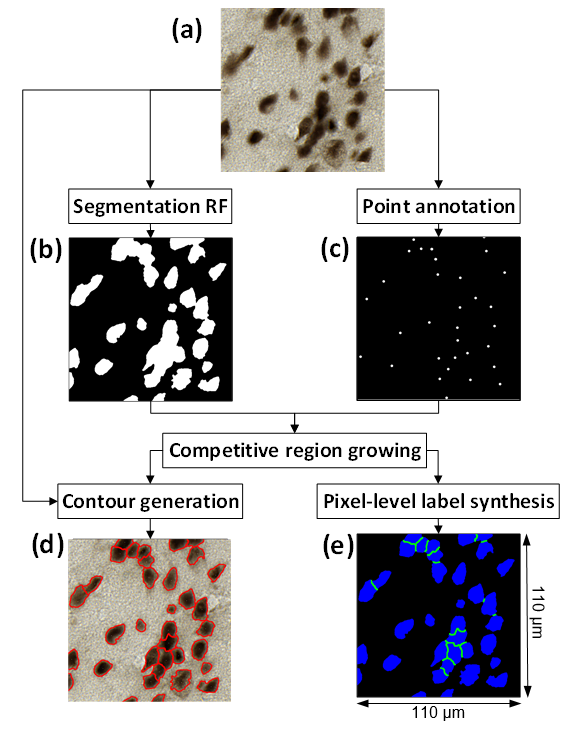}
  \caption{Flowchart of pixel-level label synthesis: (a) raw image, (b) RF segmentation, (c) point labels, (d) region growing result (red contours) and (e) synthesized masks (black: background, blue: neurons and green: inter-cell contours).}
  \label{fig:fig2}
\end{figure}

Traditionally, cell instance segmentation is addressed as a binary classification problem, the output of the classifier contains two classes: cells and tissue background. Recent works indicate that a more accurate segmentation can be achieved by classifying pixels into three classes: inter-cell contours, interior of cells and background~\cite{caicedo2019nucleus}. We designed an algorithm for constructing pixel-level labels including these three classes, based on centroid point labels provided by the expert for each neuron and binary semantic segmentation results produced in previous works~\cite{you2019automated} (as shown in Figure \ref{fig:fig2}. i) The binary segmentation of neurons and background was generated by applying a RF model of 100 decision trees, which was trained and optimized with following features: H, S, V color channels and local intensity~\cite{you2019automated}. ii) Manually pinpointed centroid labels were identified by an expert, a disk with a radius of 5 pixels in the center of the neuron was marked as a point label (size and form adapted to visual checking and region growing initialization). iii) A competitive region growing process was applied to define pixel-level labels, the expansion of sub-region was constrained using the segmentation result of RF. Each pixel inside the neuron had the same label as its centroid. By applying the contours generated by region growing on raw image Figure \ref{fig:fig2} (d), we visually assessed the quality of synthesized labels; iv) based on the pixel-level labels, three-classes masks were generated using morphological operations, including background, neurons and the border of touching neurons (thickness of 4 pixels), which is proved to enhance the segmentation result~\cite{caicedo2019nucleus}.

\subsection{Neural network architecture}
We applied the winner algorithm of the 2018 Data Science Bowl~\cite{caicedo2019nucleus}, an ensemble model of 8 U-Net-like encoder-decoder architectures, with encoders pretrained on ImageNet database, including three ResNets (34, 101, 152)~\cite{he2016deep}, two Dual Path Networks (DPN) 92~\cite{chen2017dual}, two DenseNets (121, 169)~\cite{huang2017densely} and one Inception-ResNet~\cite{szegedy2017inception}. Such strategy also enabled us to compare the performance of the different neural networks individually and to evaluate the effectiveness of the entire model. Pretrained model derived from DSB (Topcoders\_bowl)~\cite{caicedo2019nucleus}, model trained on the neuron dataset (Topcoders\_neuron) and its constituents (8 models) were all tested. Training dataset was randomly divided into two groups ($ \frac{3}{4} $ and $ \frac{1}{4} $) for training and validation respectively. A heavy data augmentation was applied to prevent over-fitting, including rotation, flipping, channel shuffling, color inversion, etc. The training set was expanded to 6 times compared to its original size. Once the training of DL models was accomplished, a post-processing step aiming to optimize the segmentation results was applied: a regression model (gradient-boosted trees) was trained to predict Intersection-over-Union (IoU) for all cell candidates, candidates with small predicted I oU ($<$ 0.3) were removed in order to decrease false predictions. 

\subsection{Overlapping extraction \& assembling}
The test dataset contained 6 large high-resolution images (5000 $\times$ 5000 pixels) which included various anatomical regions. However, most DL based segmentation algorithms cannot process such large-scale images due to GPU RAM limitation. Moreover, one constraint of CNN is that the prediction at the border of the input image is not accurate due to the lack of context information. To address this problem, we adopted the strategy of overlapped patch extraction and assembling proposed in [4]. The patches were extracted from raw images by a sliding window of 1340 $\times$ 1340 pixels (determined according to the GPU RAM resources), a stride of 1220 pixels in height and width produced an overlap of 120 pixels. The prediction results of patches were seamlessly stitched to reconstruct the final result (5000 $\times$ 5000 pixels) using the same settings as well as a weight map, which was applied to each predicted patch so that the pixels closer to the edge of the patch have lower weights. The use of the weight map reduced the impact of inaccurate prediction of pixels at the border of the patches.

\subsection{Evaluation metrics}
To evaluate the proposed method, we computed F1 score (F) for both detection (det-F1) and instance segmentation (seg-F1) tasks:

\begin{equation}
P=\frac{TP}{TP+FP}; R=\frac{TP}{TP+FN}; F=2\frac{P\times R}{P+R}
\end{equation}

where True Positive (TP), False Positive (FP) and False Negative (FN) represent the numbers of true, false and missing detection/instance segmentation respectively. For the detection task, a neuron detected was considered as a true positive when it was superimposed with exactly one centroid defined by the expert. As for the instance segmentation, the true positive was defined as the IoU greater than 0.5 between the detected neuron and the corresponding label. 

Dice coefficient was calculated to evaluate the semantic segmentation. Another evaluation criterion was the relative count error (RCE):

\begin{equation}
\varepsilon =\frac{\left | N_{a}-N_{e} \right |}{N_{e}}
\end{equation}

Where N\_a, N\_e are the number of neurons detected by the proposed method and the expert respectively.

\subsection{Training details}
Models were trained using PyTorch, Keras and Tensorflow frameworks. Each model needed different epoch number to converge (from 17 to 70 depending on network depth), with Adam optimizer and a starting learning rate of 1e-4 which decreased during the training. DL models with ResNet101, ResNet152 and one of DPN encoders used sigmoid activation, the other models used softmax activation. For loss calculation, the combination of soft dice and binary cross-entropy/categorical cross-entropy was chosen for sigmoid/softmax activation respectively.

This work was conducted on a workstation equipped with bi-processors (operating system: Ubuntu 16.04 LTS 64-bits, CPU: Intel Xeon E5-2630 v3 at 2.4 GHz, RAM: 128 GB, GPU: NVIDIA GTX 1080Ti).

\section{Results}
The results of Topcoders\_bowl, Topcoders\_neuron and its constitutive models (named according to the encoder network) were evaluated on a test dataset including $\sim$16k neurons. Table \ref{tab:table} reports the performance of neuron detection (det-F1), instance segmentation (seg-F1), semantic segmentation (Dice) and neuron counting (RCE). 
Although the training dataset of Topcoders\_bowl did not include neuron data, it was able to detect most neurons correctly (det-F1 score of 0.83 and RCE of 0.22). Nevertheless, it performed less well in both instance (seg-F1: 0.71) and semantic (Dice: 0.75) segmentation. A significant improvement of 10\% was obtained by training with NeuN stained data. Topcoders\_neuron achieved the highest detection accuracy (det-F1: 0.927), it was also one of the best models for instance segmentation (seg-F1: 0.87). Among the constitutive models, ResNet101 outperformed others in neuron detection (det-F1: 0.926) and counting (RCE: 0.037). DPN softmax was the best model for instance and semantic segmentation (seg-F1: 0.88 and Dice: 0.95), followed by ResNet34 (seg-F1: 0.87 and Dice: 0.93). 

\begin{table}
 \caption{Detection and segmentation performance of Topcoder\_bowl, Topcoder\_neuron and its constitutive models. Best and second best results are in bold with the best also underlined.}
  \centering
  \begin{tabular}{llllll}
    \toprule
    \multicolumn{2}{c}{Model}	&	det-F1	&	seg-F1	&	Dice	&	RCE	\\
    \midrule
    \multicolumn{2}{c}{Topcoder\_bowl}     &	0.825	& 0.705	&	0.751	&	0.219	\\
    \midrule
    \multicolumn{2}{c}{Topcoder\_neuron}     &	\underline{\textbf{0.927}}	& \textbf{0.870}	&	0.928	&	0.040	\\
    \midrule
    \multirow{8}{*}{\rotatebox[origin=c]{90}{Constituents}}	&	DenseNet121	&	0.910	&	0.783	&	0.830	&	0.063	\\
     & DenseNet169	&	0.912	&	0.839	&	0.869	&	0.062	\\
     & DPN sigmoid	&	0.869	&	0.803	&	0.885	&	0.130	\\
     & DPN softmax	&	0.918	&	\underline{\textbf{0.880}}	&	\underline{\textbf{0.951}}	&	0.059	\\
     & Inception-ResNet	&	0.901	&	0.848	&	0.910	&	0.116\\
     & ResNet34	&	0.914	&	\textbf{0.870}	&	\textbf{0.934}	&	0.088	\\
     & ResNet101	&	\textbf{0.926}	&	0.868	&	0.921	&	\underline{\textbf{0.037}}	\\
     & ResNet152	&	0.923	&	0.865	&	0.923	&	\textbf{0.038}	\\
    \bottomrule
  \end{tabular}
  \label{tab:table}
\end{table}

\begin{figure}
  \centering
  \includegraphics[width=0.7\textwidth]{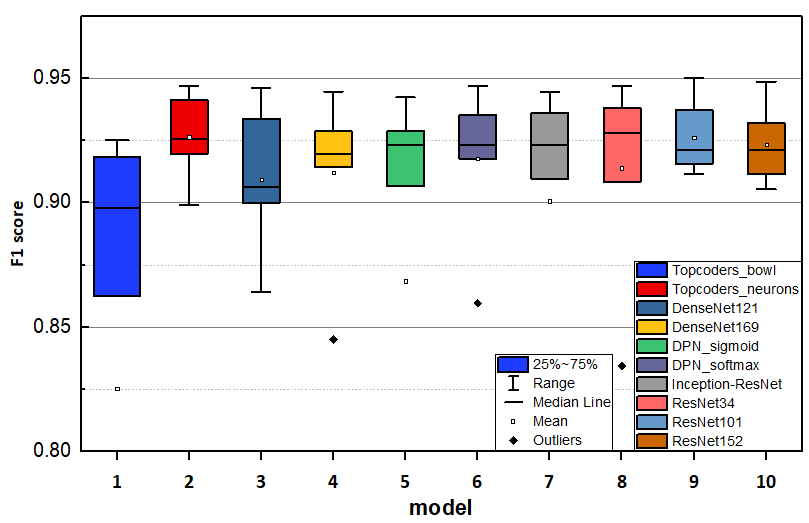}
  \caption{Comparison of detection performance (det-F1) on the test set (several outlier values are below the vertical scale range).}
  \label{fig:fig3}
\end{figure}

To better compare the performance of each model, the detection accuracy (det-F1) distribution of the 10 models are plotted in Figure \ref{fig:fig3}.  Topcoders\_neuron, which had the best average detection accuracy on the test dataset, was also one of the most robust models, it performed well for all tested anatomical regions. The best constitutive model, ResNet101, was also robust, whilst it performed less well for most tested anatomical regions than Topcoders\_neuron.

\begin{figure}
  \centering
  \includegraphics[width=\textwidth]{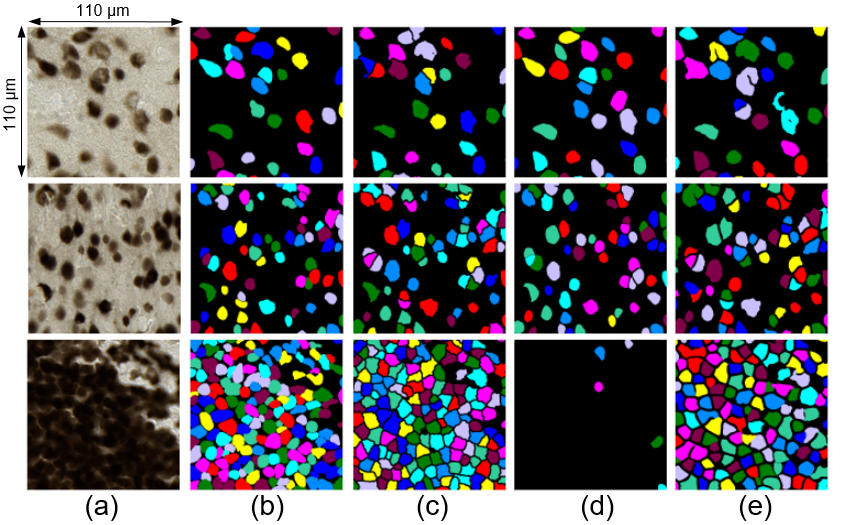}
  \caption{Results of pixel-level label synthesis and segmentation in three anatomical regions. Top row: caudate; Middle row: cortex; Bottom row: hippocampus. (a) Raw images, (b) manual segmentation, (c) synthesized pixel-level masks and (d, e) results of Topcoders\_bowl and Topcoders\_neuron respectively.}
  \label{fig:fig4}
\end{figure}

Figure \ref{fig:fig4} shows cropped examples of synthesized masks and segmentation results in three anatomical regions. Neurons are presented by colored labels. The results are displayed according to an increasing density of neurons (first row: caudate, sparse; second row: cortex, dense and last row: hippocampus, very dense). For better illustrating the results, the neurons in these regions were segmented manually, as shown in Figure \ref{fig:fig4} (b). Figure \ref{fig:fig4} (c) shows the synthesized pixel-level masks, compared to the manual annotations, the proposed method produced satisfying masks for most regions (both in distribution and shape based on visual evaluation). Figure \ref{fig:fig4} (d) illustrates the results of Topcoders\_bowl, the neurons were correctly detected in the sparse regions, but the contour of neurons were often distorted. Moreover, the region of massive touching neurons was wrongly considered as the background. Figure \ref{fig:fig4} (e) presents the predictions of Topcoders\_neuron, it was a solid model performing very well for all tested anatomical regions, at both object-level and pixel-level.

\section{Discussion}
Topcoders\_bowl was believed to be a well generalized model based on DSB 2018 results. It successfully separated neurons in regions with sparse distribution while it did not respect the neuron shape and it failed in regions where massive neurons aggregated. This is probably due to the absence of cells with various forms and highly clustered distribution such as neurons in the dataset of DSB. 

The performance of Topcoders\_neuron demonstrated the superiority of the ensemble model. Although the results of Topcoders\_neuron were obtained by combining the predictions of the constitutive models, it achieved better results than most constituents for all the tasks. 

Among all the constitutive models, all ResNet backbone models performed well, including ResNet34, the best model for semantic and instance segmentation, and ResNet101, the best constitutive model for neuron detection. Generally, a deeper network can capture more complex features. While the deepest model ResNet152 did not achieve the best performance in any task, it might be related to the fact that deeper networks are generally more difficult to train owning to the vanishing gradient problem~\cite{tan2019efficientnet}. Another interesting finding is that the choice of activation had an important influence on segmentation results. Although two DPN models had exactly the same architecture, the one with softmax activation performed better than that with sigmoid activation. 
Since most tested neural networks achieved good results, we believe that it is feasible to apply DL techniques for neuron counting. However, compared to stereology~\cite{west1991unbiased}, which takes into account the thickness of tissue for possible superimposition of cells and provides unbiased quantifications, DL methods can only deal with 2D images and provide a valuable estimation of cell counting. A specific dataset and study need to be designed to quantify discrepancies between DL methods and stereology.

\section{Conclusion}
In this work, we investigated the ability of a weakly supervised method to specifically detect and segment neurons in NeuN stained histology images. By applying state-of-the-art DL architectures, this study provides the first comprehensive assessment of different neural networks for neuron individualization. An optimal model trained using neuron data was obtained, it was able to separate size, shape and density-varying neurons successfully. Experimental results in the main anatomical regions demonstrated the effectiveness of the proposed method against the default DSB model. The current study was carried out with the default settings, further optimization in training parameters and architecture need to be investigated. Besides, developments in high-performance computing (HPC)$\dagger$ are also ongoing to test the efficiency of cross-validation. Further work is required to establish a comparative analysis of Topcoders and other deep learning-based instance segmentation methods, as well as stereology – the reference method used in biomedical analysis. An exciting perspective will be to extend this study to whole sections and brains, which will improve our understanding of brain development, aging and neurodegeneration.

\section*{Compliance With Ethical Standards}
The experimental procedures involving animal models described in this paper were approved by the Institutional Animal Care and Ethics Committee.

\section*{Acknowledgment}
$\dagger$ This work was granted access to the HPC resources of TGCC under the allocation 2019-(A0040310374) made by the GENCI.

\bibliographystyle{unsrt}  
\bibliography{references}

\end{document}